\documentclass{sig-alternate-05-2015}
\newcommand{\minitab}[2][l]{\begin{tabular}{#1}#2\end{tabular}} 

\usepackage{amssymb}
\usepackage{xspace}
\newcommand{\etal}{{\em et al.}\xspace}
\usepackage{multirow} 
\usepackage{subfigure} 
\usepackage{array} 
\usepackage{url} 
\usepackage{graphicx}
\begin{document}
\title{Andro-profiler: Detecting and Classifying Android Malware based on Behavioral Profiles}
\author{Jae-wook Jang \\ Korea University \and Jaesung Yun \\Korea University \and  Aziz Mohaisen \\  University at Buffalo \and Jiyoung Woo \\ Korea University \and Huy Kang Kim \\ Korea University}
\maketitle
\begin{abstract} 
Mass-market mobile security threats have increased recently due to the growth of mobile technologies and the popularity of mobile devices. Accordingly, techniques have been introduced for identifying, classifying, and defending against mobile threats utilizing static,  dynamic, on-device, off-device, and hybrid approaches.
In this paper, we contribute to the mobile security defense posture by introducing Andro-profiler, a hybrid behavior based analysis and classification system for mobile malware. Andro-profiler classifies malware by exploiting the behavior profiling extracted from the integrated system logs including system calls, which are implicitly equivalent to distinct behavior characteristics. Andro-profiler executes a malicious application on an emulator in order to generate the integrated system logs, and creates human-readable behavior profiles by analyzing the integrated system logs. By comparing the behavior profile of malicious application with representative behavior profile for each malware family, Andro-profiler detects and classifies it into malware families. The experiment results demonstrate that Andro-profiler is scalable, performs well in detecting and classifying malware with accuracy greater than $98\%$, outperforms the existing state-of-the-art work, and is capable of identifying zero-day mobile malware samples.
\end{abstract}

{\noindent\bf Keywords.} 
Behavior profiling, Similarity, System call, Android, Malware


%




\section{Introduction}

The explosive growth in the number of mobile devices running the Android platform has attracted the attention of hackers for the wealth of sensitive information that are usually stored on mobile devices, including phone numbers, short messages, confidential emails and correspondences, and banking information and credentials. The availability of this information in many mass-market mobile devices makes them a desirable target for hackers, who excelled at developing a large number of mobile malicious software (malware), making the security of mobile devices one of the most important and challenging areas of research.
For example, According to a report by McAfee, the total number of mobile malware continued its linear climb as it broke 8 million in the second quarter of 2015, and increased by 17\% over the first quarter of the same year \cite{McAfee2015}. Moreover, new malware families and variants were reported to appear approximately 1 million times in the same quarter.
To address this trend, antivirus (AV) vendors analyze a large number of malware samples daily in order to prevent them from spreading widely and to guide users on disinfection and risk management by classifying malware into broad families.

Mobile as well as traditional malware analysis for detection and classification falls into two broad types: static and dynamic analysis. In static analysis, strings of bytes associated with malware samples are discovered through reverse engineering and used as a signature for identifying malicious software. Although fast and efficient, static techniques are often prone to high false positive rates due to evolution in code basis and code repacking. Furthermore, additional cost of those techniques is required for reverse engineering to generate reliable and meaningful signatures.

On the other hand, dynamic and behavior based analysis aims to provide methods for effectively and efficiently extracting unique patterns of each malware family based on its behavior. Malware samples of the same family often use the same code base, provide the same functionality using the same order of behavioral events~\cite{MohaisenWMA14}, and so on. In analyzing mobile malware, unique behavior patterns can be represented by various symbols (e.g., permission set, API call, and system call) and used to identify malware families. To this end, researchers previously proposed various detection and classification methods for malware analysis based on their behavior, including permission-based, API call-based and system call-based methods. Permission-based detection methods are not efficient in classifying benign applications as benign, since relevant rule sets only focus on detecting the malware. API call-based detection methods cannot generate distinct signatures until  decompilation or disassembly process is completed, which is often expensive. System call-based detection methods can more accurately detect malicious behavior than other methods, since it is impossible to modify original functionality of system calls: malware creators always attempt to disguise malicious behavior as normal behavior. However, proposed methods in this category mainly deal with frequency of system calls well presented in malware. The number of invoked system calls is usually small, and most of the system calls used in malware (e.g., {\tt read()}, {\tt write()}) are also observed in both benign applications, affecting the  accuracy of those methods. To this end, one needs to consider more features, such as arguments in the system call and network activities, to enhance malware detection and classification via behavior profiling.

To overcome the drawbacks in previous methods, we propose a feature-rich anti-malware system based on behavior profiling called Andro-profiler. Our proposed behavior profiling system comprises mobile devices and a remote server to facilitate profiling, and adopts profiling method in the malware analysis domain.
We exploit system calls, including their arguments provided by LKM (Loadable Kernel Module) and system logs (e.g., SMS, call, and network I/O) provided by Droidbox \cite{Droidbox} as feature vectors for malware characterization. We define system calls and system logs as integrated system logs from which we directly infer behavior patterns representation using the concept of behavior profiling of Bayer \cite{Bayer2009}.
We assume that: a) malware samples have unique malicious behavior patterns, b) malicious behavior is determined by system calls, and c) such system call set has influence on the behavior of the program (malware). We prepare representative behavior profile for each malware family represented by integrated system logs including system calls, their arguments, and system logs of Droidbox--an analysis system we utilize in this work. We construct the behavior profile of each malware sample through its integrated system logs by executing it on an emulator. Then, by comparing the behavior profiles  across samples, we can detect and classify malware samples into related families.

\vspace{3mm}
{\noindent\bf Contribution:} The main contributions of this paper are as follows:
\begin{enumerate}
\item We propose a novel anti-malware system based on behavior profiling called Andro-profiler. We classify malware by exploiting the behavior profiling extracted from integrated system logs. Our method captures the behavior profiling by converting integrated system logs into human-readable contexts, which helps analysts analyze malware intuitively.

\item Andro-profiler enables AV vendors to react to many species of malicious samples by classifying and matching them with those previously detected quickly and efficiently. Our system can help detect new malware including existing malware's variants and zero-day exploits. This is further highlighted through in-depth experiments using real-world malware samples.

\item Our proposed method is robust, and can be extended with additional features that depict the unique behavior patterns of malware. Our method can easily employ static analysis technique to capture malicious behavior, in combination of the dynamic behavior, which is shown to outperform existing techniques in the literature. This feature of our work is highlighted by a comparison with the prior literature, experimentally.
\end{enumerate}


\section{Related Work}

Based on where the scan and monitoring of the mobile malware takes place, malware analysis methods are classified into three types: detection methods on the mobile device, detection methods outside the mobile device, and hybrid detection methods. We classify the literature based on the type of the malicious behavior into permission-based and footprint-based methods. Footprint-based methods include system call-based, API call-based, decompiled code-based, and XML information-based methods. The detection methods on mobile device scan malicious behavior patterns on the mobile device and return the analysis results to the user. However, those approaches do not consider the resource constraints on the mobile device: low computing power and limited battery life, affect their usability and user experience. The detection methods outside mobile device execute detection algorithms on an emulator or a real device running the targeted applications, and conduct static or dynamic analysis for determining the nature of those applications. Those approaches do not need to consider resource constraints, but cannot respond to new malware families quickly. To overcome the drawbacks in both approaches, hybrid approaches have been introduced in mobile malware analysis. Client modules deployed on mobile devices collect information related to installed applications on those devices and send the information to a remote server. The remote server then analyzes log files using their detection algorithms of choice, while not impeding usability and user experience.
 Table ~\ref{tab:related} summarizes the various malware detection or classification methods in the literature. In the following, we elaborate on some of the related works in each category.

\begin{table}[h!]
\centering
\renewcommand{\arraystretch}{1.3}
{\tiny
\caption{Various malware detection$/$classification methods in previous works.}
\label{tab:related}
\setlength{\tabcolsep}{1pt}
\begin{tabular}{c c c c}

\hline
Approach & Method & Feature & Previous works\\
\hline
\multirow{5}{*}{\begin{tabular}{@{}c@{}}Detection on \\ mobile device\end{tabular}} & Permission & Permission & \begin{tabular}{@{}c@{}}\cite{Enck2009} \\ \cite{Pearce2012} \end{tabular}\\
\cline{2-4}
& \multirow{3}{*}{Footprint} & System resources & \begin{tabular}{@{}c@{}}\ \cite{Shabtai2010} \\ \cite{Bugiel2012} \end{tabular} \\
\cline{3-4}
& & Taint tracing & \cite{Enck2010} \\
\cline{3-4}
& & Event log, System call & \cite{Bose2008} \\
\hline
\multirow{12}{*}{\begin{tabular}{@{}c@{}}Detection \\ outside \\ mobile device \end{tabular} } & Permission & Permission & \cite{Peng2012} \\
\cline{2-4}
& \multirow{3}{*}{Footprint} & System call, Disassembled code & \cite{Blasing2010} \\
\cline{3-4}
&  & System call, Interaction log & \cite{reina2013} \\
\cline{3-4}
&  & System/API call, Taint tracing & \cite{Rastogi2013} \\
\cline{2-4}
& \multirow{8}{*}{\begin{tabular}{@{}c@{}}Permission + \\ Footprint \end{tabular}  }  & Permission, API call & \cite{Yang2012} \\
\cline{3-4}
& & \begin{tabular}{@{}c@{}} Permission, API call, \\ XML information \end{tabular} & \begin{tabular}{@{}c@{}}  \cite{Grace2012} \\ \cite{Wu2012} \\ \cite{Arp2014} \\ \end{tabular} \\
\cline{3-4}
& & \begin{tabular}{@{}c@{}}Permission, API call, \\ System call, \\ XML information, \\ Disassembled code \end{tabular}  &  \begin{tabular}{@{}c@{}} \cite{Yan2012} \\ \cite{Zhou2012} \\ \cite{Spreitzenbarth2013} \end{tabular} \\
\cline{2-4}
\hline
\multirow{2}{*}{Hybrid} & \multirow{2}{*}{Footprint} & System call & \begin{tabular}{@{}c@{}}\cite{Burguera2011} \\ \cite{Isohara2011}  \end{tabular}\\
\cline{3-4}
& & Function call & \cite{Schmidt2009} \\
\hline
\end{tabular}
}
\end{table}

\subsection{Detection Methods on Mobile Devices}

Previous work in this category has introduced malware detection methods that can execute applications on devices, providing online detection.
Enck \etal \cite{Enck2009} proposed the Kirin security service, which performs lightweight certification of applications to mitigate malware at installation time. Kirin examined the requested permissions of applications, compared them with self-defined security rules, and determined whether malicious activities were carried out or not. In order to do that, they  relied on permissions given in a manifest file, {\tt Androidmanifest.xml}. However, application developers tend to excessively declare permissions in a manifest file, although the application does not actually need all of the permissions. To that end, those methods produce low accuracy.
Pearce \etal \cite{Pearce2012} introduced AdDroid, in which they separated advertising permissions for the Android platform. In AdDroid, the host application and the core advertising code ran in isolated environment, where applications using AdDroid did not send sensitive information to advertisement server anymore. However, AdDroid did not consider information leakage unrelated to advertisement, which is the case in the majority of malware.

Shabtai \etal \cite{Shabtai2010} proposed Andromaly, a behavior-based detection framework for Android-based mobile devices. Andromaly is a host-based intrusion detection system that continuously monitored various resources and classified malicious applications using a machine learning algorithm. These proposed methods, however, require a significant hardware capacity (e.g., CPU, RAM, and battery life) in order to monitor all resources comprehensively.
Bugiel \etal \cite{Bugiel2012} proposed Xmandroid, a system-centric and policy-driven runtime monitoring system that regulates communications between applications. Based on heuristic analysis, the authors identified attack patterns and classified malicious applications.

Enck \etal \cite{Enck2010} proposed Taintdroid, an extension to the Android mobile-phone platform that tracks the flow of sensitive information through third-party applications. If tainted data left the Android device, Taintdroid provided a report logging the leaked data, where the data is sent and which application leaked it. Taintdroid focused on information leakage, and then an emulator such as Droidbox embedded Taintdroid and tracked information leakage.

Bose \etal \cite{Bose2008} proposed a signature-based detection method for the Symbian operating system. The method is a two-stage mapping technique consisting of extraction process and representation process that constructed these signatures at run-time from the monitored system events and system calls. The method used temporal logic to detect malicious activity over time that matched a set of signatures represented as a sequence of events. However, the method needed to obtain root privileges to access the kernel, and required sufficient hardware capacity to extract system calls and convert related features into signatures.

\subsection{Detection Methods Outside Mobile Devices}

Previous work in this category introduced malware detection methods that execute relevant applications outside the device, providing offline detection. These methods execute their detection algorithms on an emulator or a real device other than the host device. Thus they are not constrained by constraints of real devices, and do not impede usability and user experience.

Peng \etal \cite{Peng2012} used probabilistic generative models for risk scoring schemes, ranging from the simple Na\"ive Bayes to advanced hierarchical mixture models. Their proposed methods computed a real risk score of Android applications based on the requested permissions, and differentiated between malware and benign applications. However, application developers tend to excessively declare permissions in a manifest file, requiring the method to rely on other criteria for higher detection and classification accuracy.

Blasing \etal \cite{Blasing2010} proposed an Android Application Sandbox (AASandbox), which enables static and dynamic analysis on the Android platform. In the static analysis phase, AASandbox decompressed installation files and disassembled intended executable files, then compared them with pre-defined malicious patterns. In the dynamic analysis phase, it hijacked system calls for logging and built a frequency table of system calls. However, the dynamic analysis methods based on the frequency of system calls need a more elaborate and redefined process in order to improve its detection or classification accuracy. The function name of the system call as well as arguments used in the system call need to be considered.
Reina \etal \cite{reina2013} introduced CopperDroid, an approach built on top of QEMU to automatically perform dynamic analysis of Android malware. CopperDroid conducted a unified analysis to characterize low-level OS-specific and high-level Android-specific behaviors (e.g., information leakage, sending SMS) by observing and analyzing system call invocations, IPC and RPC interactions.
Rastogi \etal \cite{Rastogi2013} proposed AppsPlayground, a framework for automatic dynamic analysis, which executes a suspicious application on emulator built on top of QEMU. AppsPlayground determined whether malicious activities were carried out or not by tracking information leakage and monitoring sensitive API and system calls.

Yang \etal \cite{Yang2012} introduced a systematic approach, called Money-Guard, to detect stealthy money-stealing applications in the Android market. Money-Guard checked for API calls and billing-related permissions to detect stealthy money-stealing malware, but could not identify various malicious behavioral patterns except for malware sending premium-rate SMS.
Grace \etal \cite{Grace2012} proposed an anti-malware system, RiskRanker, to determine whether or not an application conducts malicious behavior by measuring potential security risk. RiskRanker classified an application into a high-risk application if it had exploit code for vulnerabilities in the OS. RiskRanker reported an application as a medium-risk application that enables to hijack sensitive information or subscribe premium service without victim's consent. Moreover, RiskRanker inspects malware embedding encryption and dynamic loading methods.
Wu \etal \cite{Wu2012} proposed DroidMat, which is a feature based malware detection method. DroidMat chose the requested permissions, Intent message, and API calls as feature vectors, extracted them from various resources such as manifest file and bytecode. By leveraging a $K$-means clustering algorithm, DroidMat modeled malware samples according to their characteristics, and then determine whether or not an application is malicious by leveraging $K$-NN algorithm.
Arp \etal \cite{Arp2014} proposed DREBIN which utilizes the used permissions, suspicious API calls, and network addresses as feature vectors for identifying applications. DREBIN extracted those features from the manifest and {\tt dex} bytecode files, and identified malware by leveraging Support Vector Machine (SVM) algorithm.

Yan \etal \cite{Yan2012} proposed DroidScope built on top of QEMU and enabled to reconstruct the OS-level and Java-level semantic views simultaneously. They analyzed malware by collecting native and Dalvik instruction traces, API-level activity, and information leakage. Zhou \etal \cite{Zhou2012} proposed DroidRanger, which identifies malicious behavior through both permission-based behavioral footprint scheme for the detection of known malware and a heuristic-based filtering scheme for detection of zero-day malware.
Spreitzenbarth \etal \cite{Spreitzenbarth2013} proposed Mobile-Sandbox, static and dynamic analyzer for Android applications, like in AASandbox. In the static analysis phase, Mobile-Sandbox parsed a manifest file, decompiled the application, and checked whether suspicious permissions are used or not. In the dynamic analysis phase, they executed the application on Droidbox, logged every operation of the application, and recorded native library calls executed by processes. They extracted native library calls by exploiting ltrace \cite{ltrace}; ltrace is executed after installation process is completed.

\subsection{Hybrid Methods}

In hybrid detection methods, clients collect meta information on applications on the device and send that information to a remote server. The remote server then analyzes this information using a detection algorithm and makes a decision on whether an application is benign or malicious. This approach compensates for the drawbacks of the online and offline detection methods. However, users have to agree in advance on what client module will send user information to the remote server.

Burguera \etal \cite{Burguera2011} proposed a lightweight client called Crowdroid which monitored system calls, made a frequency table using those system calls, and sent them to a centralized server. The remote server then identified malicious behavior in a statistical manner and detected malware using a $K$-means clustering algorithm. Crowdroid extracted system calls by exploiting Strace \cite{Strace}, but Strace is executed after installation---Crowdroid cannot detect malicious behavior during the installation process, and depends on the functionality of Strace.
Isohara \etal \cite{Isohara2011} proposed a kernel-based behavior analysis system that consisted of a system call log collector on an Android device and a log analyzer on a remote server. The client collected system calls generated at installation time and sent the logs to a remote server. The remote server then compared patterns in the logs with 16 pre-defined patterns. Since pre-defined behavior patterns mainly focused on malicious behaviors such as restricted information leakage, jailbreak, and abuse of root privileges, their system could not detect malicious behavior such as sending premium-rate SMS and calling premium-rate code. They also do not guarantee sufficient scalability.

Schmidt \etal \cite{Schmidt2009} proposed a collaboration mechanism for Android platform security comprising a log collector on the device and a remote analyzer. In their proposed system, the client monitored the behavior of the malicious application at the installation time, ran analysis based on the similarity of the function call set used, exchanged the result of analysis with neighboring devices, and performed collaborative malware detection.

{\bf\noindent Other methods.} Other methods that look at mal-actor specific information for detection and classification have been explored as well. For example, Jang \etal \cite{jang2015andro} proposed Andro-autopsy, an anti-malware system based on malware creator information, Andro-dumpsys~\cite{jang2016andro}, which utilizes malware centric information (mainly utilizing memory usage), and Mal-Netminer~\cite{jang2015mal}, which utilizes network theoretic approach to feature extraction and classification. Kang \etal introduced Andro-Tracker~\cite{kang2015detecting,kang2014androtracker}, which utilizes static android data for classification, and

\section{Behavior Profiling}\label{sec:behavior_profiling}
In the literature of traditional malware research related to personal computers operating Microsoft Windows,  Bayer \etal\cite{Bayer2009} proposed a method for scalable behavior-based malware clustering. The method contributes to the theoretical foundations of malware analysis by discussing the behavior-based profiling formally. Given the relevance of this work to our work, we review definitions of behavior profiling from the aforementioned work for the completeness of our presentation, and incorporate details specific to our proposed system in the following.

\vspace {0.2 cm}
\textbf{Definition} (behavior profiling). A behavior profiling $P$ is defined by four tuples as $P = (O, OP, \Gamma, \Delta)$, where $O$ is the set of all objects and $OP$ is the set of all operations, which is represented in nested dictionary form as \{name : \{target : attribute\}\}. $\Gamma \subseteq (O \times OP)$ is a relation assigning more than one operation to each other, and
$\Delta \subseteq ((O \times OP), (O \times OP))$ represents the sequence-unrelated set, which is equivalent to integrated system logs.

\vspace {0.2 cm}
\textbf{Object}: An object represents an abstract functionality that malware samples need for carrying out the malicious behavior.
\cite{Tam2015} manually analyzed many malware samples from various datasets such as contagion and Android Malware Genome Project. They classified malicious behaviors into six groups according to behavior patterns: make call, send SMS, network access, access personal information, alter filesystem, and execute external application. We also manually inspected malware samples we had, and then defined malicious behavior as outlined by \cite{Tam2015}; since some behavior patterns are not found in our dataset, we leave them out.
We define malicious behavior as the sending of premium-rate SMS, the calling of premium-rate number, the sending of sensitive information, and converting data for transmission. We do not consider malicious behavior such as privilege escalation and C \& C (Command and Control) attack since dynamic analysis methods hardly detect malware executing malicious behavior under given condition (e.g., SDK version, cellular network connection status, time, or place). We formally define object as following:

\vspace {0.2 cm}
\begin{verbatim}
    Object      ::= Object-type
    Object-type ::= Telephony | Phone | Network
\end{verbatim}

\vspace {0.2 cm}

\textbf{Operation}: An operation represents a concrete malicious behavior. Formally, an operation comprises of operation-name, operation-target, and operation-attribute.
Operation-name is the identifier for malicious behavior. Operation-target is the attack objective of malware, such as contents of external storage and system information. Operation-attribute is a meaningful value that the malware wants to obtain; for example, the attribute of country code (operation-target) is Korea, and operation-name is sending sensitive information. We formally define operation as follows:

{\scriptsize
\begin{verbatim}
    Operation         ::= { Operation-name : { Operation-target :
                            Operation-attribute } }

    Operation-name    ::= Sending SMS | Calling | Sending sensitive
                          information | Converting data

    Operation-target  ::= Premium-rate SMS/number | device ID |
                          IMEI | IMSI | MCC | MNC | ...| etc.
\end{verbatim}
}

Table \ref{Example} shows an example of mapping of network object and the corresponding operations. In the case of malicious behavior for sending sensitive information, we represent the profile of that behavior as follows:
 ``\{Network : \{Sending sensitive information : \{\{IMEI : 357242043237517\}, \{MCC : 310\}, \{MNC : 260\}, \{Location : GPS Coordinates \} $\cdots$, \} \} \}".

\begin{table}[h!]
\centering
\renewcommand{\arraystretch}{1.3}
{\tiny
\caption{Example of mapping of network object.}
\label{Example}
\setlength{\tabcolsep}{1pt}
\begin{tabular}{c c c c}
\hline
Type & Name & Target & Attribute \\
\hline
\multirow{14}{*}{Network} & \multirow{11}{*}{\minitab[c]{Sending \\ sensitive information}} & Android Id &  \minitab[c]{ 3531505c0b421c4d } \\
\cline{3-4}
& & Device type & Android \\
\cline{3-4}
& & IMEI & 357242043237517 \\
\cline{3-4}
& & IMSI & 310005123456789 \\
\cline{3-4}
& & MCC & 310 \\
\cline{3-4}
& & MNC & 260 \\
\cline{3-4}
& & OS version & 10 \\
\cline{3-4}
& & SDK version & 2.3.4 \\
\cline{3-4}
& & Carrier & Android \\
\cline{3-4}
& & Country code & en \\
\cline{3-4}
& & Location & GPS Coordinates \\
\cline{2-4}
& \multirow{4}{*}{Converting data} & Cipher algorithm & No, DES, AES, Blowfish \\
\cline{3-4}
& & Destination URL & \url{http://my365image.com} \\
\cline{3-4}
& & Port & 80 \\
\cline{3-4}
& & Encoding algorithm & gzip \\
\hline
\end{tabular}
}
\end{table}

\section{Andro-profiler: An Anti-Malware System}\label{sec:androprofiler}
In the following we review the design and operation of Andro-Profiler, a hybrid system for malware analysis and classification that combines the on-device capabilities for profiling and off-device capabilities for analysis and classification.

\begin{figure}[htb]
 \centering{
 \scalebox{0.25}{\includegraphics{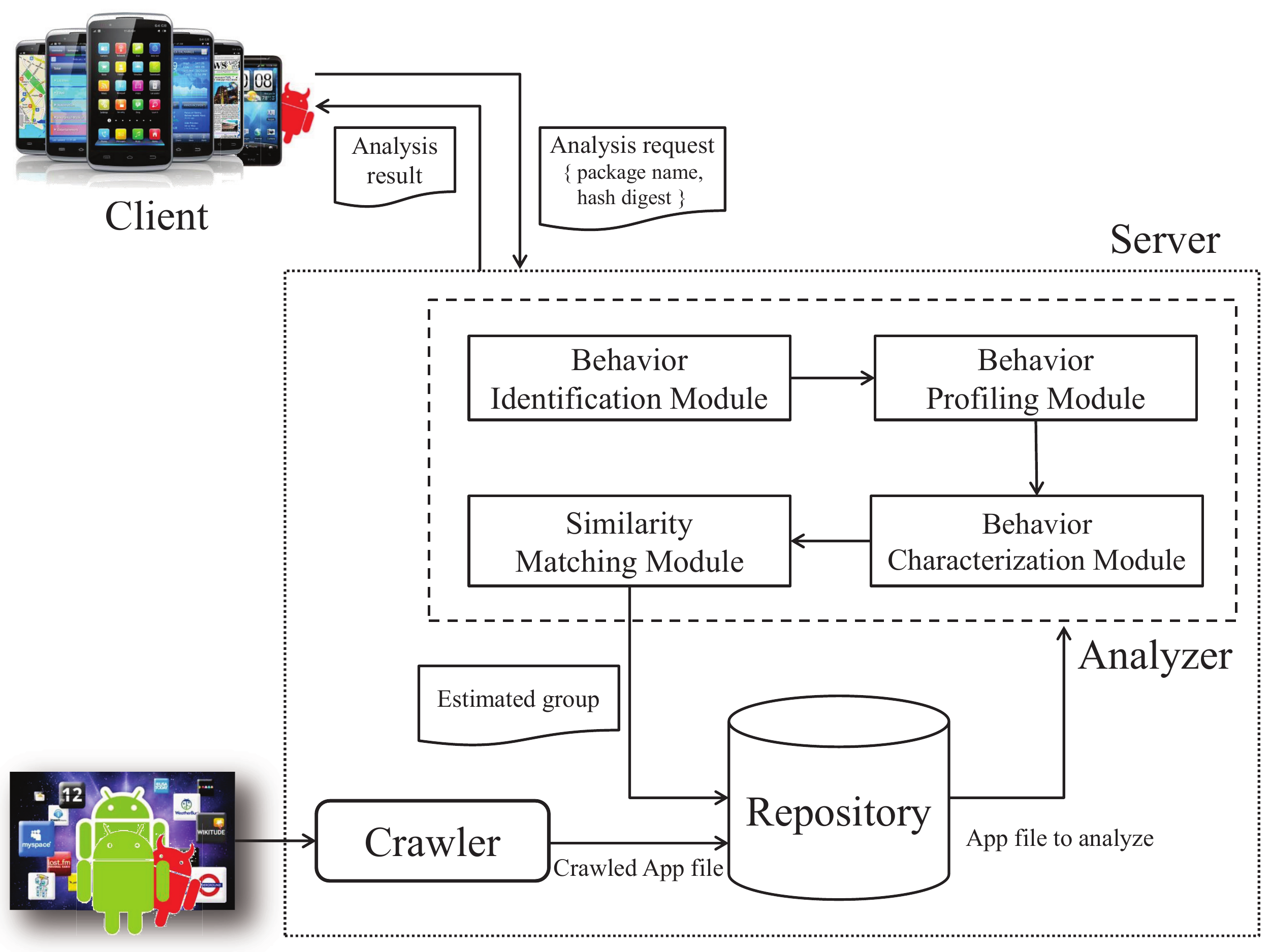}}
 \caption{Overall procedure of Andro-profiler.}
 \label{fig_1}
 }
\end{figure}

\subsection{Overview}\label{sec:3:overview}
\begin{figure}[h!]
 \centering{
 \scalebox{0.25}{\includegraphics{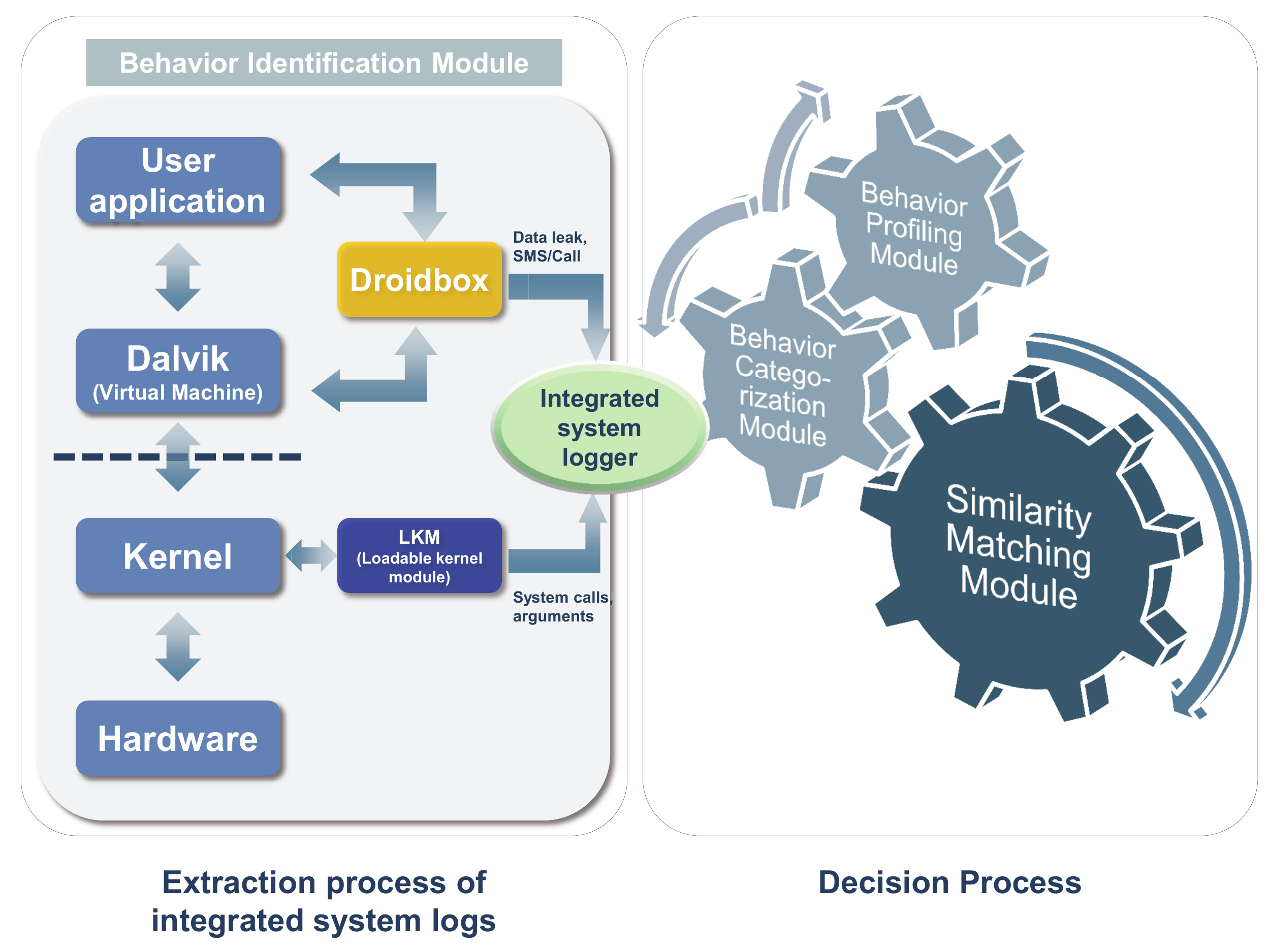}}
 \caption{{Overview of the analyzer component.}}
 \label{fig_2}
}
\end{figure}

As illustrated in Figure \ref{fig_1}, we propose a hybrid anti-malware system that consists of a client application on the mobile device and a profiling and analysis remote server. The client application on the mobile device collects installed application information, and sends that information to the remote server; the client application only sends application-specific information such as the hash digest of {\tt apk} file and package name. If the remote server cannot crawl that application, the client application sends the application package file ({\tt apk}) to the remote server. The remote server analyzes the malicious application and decides whether it is malicious or not based on its behavior. The remote server consists of three components: crawler, repository, and analyzer. The crawler component crawls applications from repositories, such as official markets and alternative markets. The crawled applications are then passed to the repository component which runs a duplication test by comparing the hash digest of the {\tt apk} file to each other. If the crawled application is a duplicate, it is discarded; otherwise, the repository component sends that application to the analyzer component. After completing the analysis, the analyzer component sends the analysis results to both the repository component and the client application. Upon receiving the analysis results from the remote server, the client application displays the result on the screen to the user.
The repository component searches its database upon the repository component receiving an analysis request from the client. If the repository component does not have analysis results to fulfill the client application's request, it fetches the crawler component. As illustrated in Figure \ref{fig_2}, the analyzer component has two processes: an extraction process of integrated system logs and a decision process. The extraction process of integrated system logs is composed of a behavior identification module, and the decision process is composed of three modules: a behavior profiling module, a behavior categorization module, and a similarity matching module. In the following, we review the extraction and decision processes.

\subsection{Extraction Process of Integrated System Logs}\label{sec:3:extraction}
 {\noindent\bf Behavior Identification module:}  Andro-profiler conducts malware characterization based on dynamic behavior analysis. Our system extended Droidbox to embed the LKM (Loadable Kernel Module) for hijacking system calls including their arguments. More specifically, the Behavior Identification (BI) module in our system executes malware on an emulator and monitors malicious behavior in an isolated environment. Whenever malware is executed on the emulator, the BI fetches the integrated system logger. The integrated system logger parses system calls including their arguments provided by LKM and system logs provided by Droidbox; Droidbox monitors SMS, call, and network I/O. The parsed integrated system logs are then passed to the {\em decision process}.

\subsection{Decision Process}\label{sec:3:decision}
As shown in Figure~\ref{fig_2}, the decision process consists of three modules: behavior profiling, behavior categorization, and similarity matching module. In the following we elaborate on each of those modules.
\subsubsection{Behavior Profiling Module}
\begin{figure*}[htb]
 \centering{
 \scalebox{0.45}{\includegraphics{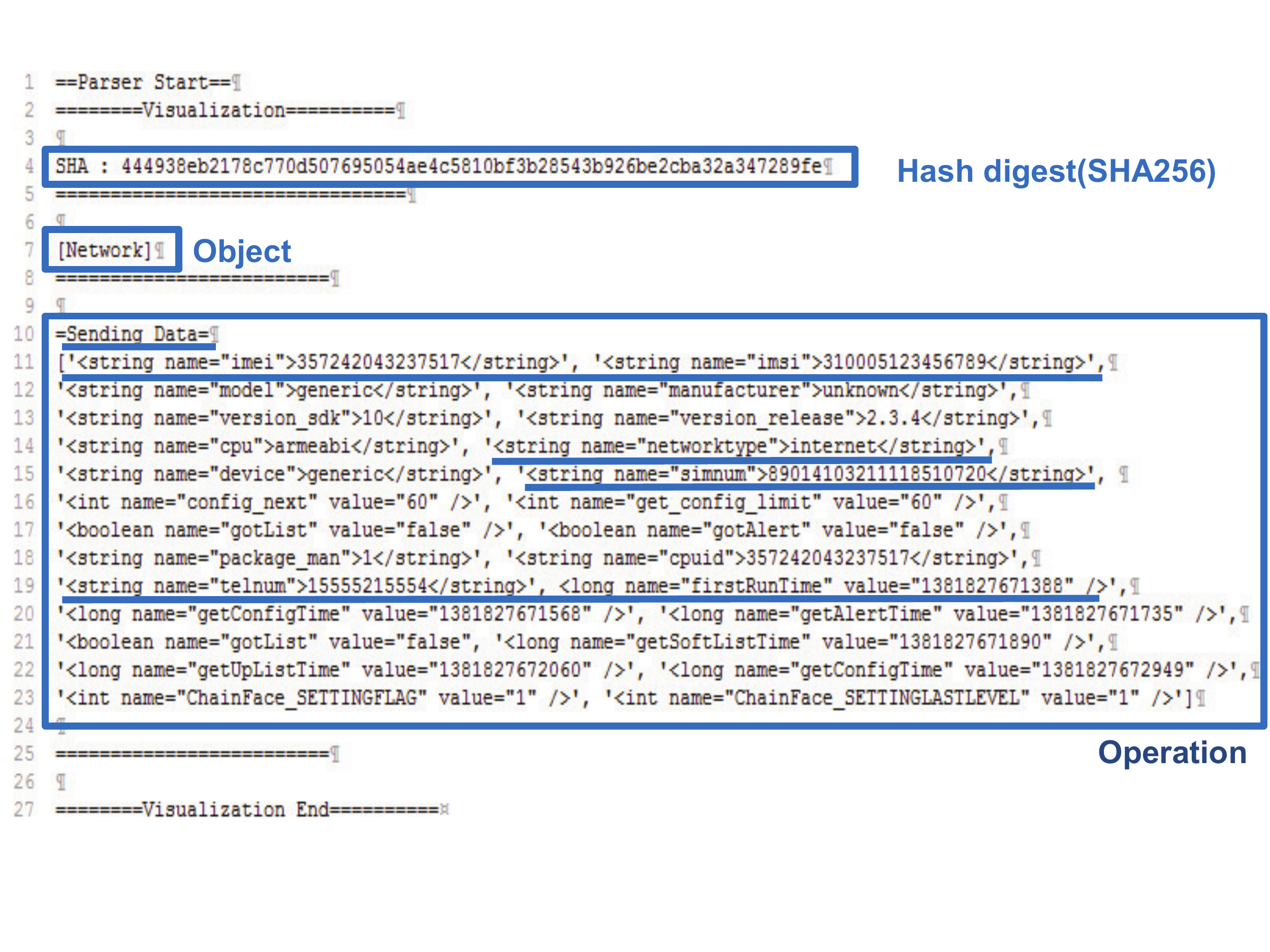}}
 \caption{{Implementation of behavior profiling (e.g., GinMaster).}}
 \label{fig_3}
 }
\end{figure*}

The Behavior Profiling (BP) module parses the integrated system logs of a given application and makes the behavior profile. The BP module is implemented as described in previous section (Behavior Profiling). For example, the BP module makes a behavior profile of {\tt GinMaster} which steals sensitive information, as illustrated in Figure \ref{fig_3}. According to the analysis report of F-Secure \cite{FSecure_Gin}, {\tt GinMaster} steals sensitive information, such as International Mobile Equipment Identity (IMEI), International Mobile Subscriber Identity (IMSI), User Identifier (UID), Subscriber Identification Module (SIM) number, telephone number, and network type, to a remote server. The behavior profile made by the BP module is similar to the analysis report of F-Secure, and it is simple and relatively easy to understand.

\subsubsection{Behavior Categorization Module} The behavior categorization (BC) module categorizes a given application according to its behavior patterns. As we mentioned earlier, we define malicious behavior as the sending of premium-rate SMS, the calling of premium-rate number, the sending of sensitive information, and converting data for transmission. Since the numbers of malicious behavior patterns which we define are four, the possible permutation sets of malicious behavior patterns are 15 $(=\sum_{i=1}^{4} {}_4C_i{})$. If an application does not behave in accordance with a pre-defined malicious behavior, our system decides that the application is benign.

\subsubsection{Similarity Matching Module} The different similarity metrics need to be applied to behavior factors since they have different types of argument. Instead of using machine learning approaches that usually use the same similarity metric for features, we design the appropriate similarity metrics for behavior factors. The similarity matching (SM) module computes the similarity score between the behavior profile of malicious application and representative behavior profile of each malware family. The SM then classifies the malicious application into the group with which it bears the most similarity based on its behavior. The representative behavior profile of each malware family has to depict the unique and common behavior patterns of each malware family, then SM module chooses one of the methods updating the representative behavior profile as follows:

\begin{enumerate}
\item {\bf Method 1}: The first update method is {\em intersection}. The representative behavior profile for each malware family is updated by the intersection of behavior profiles of members in each subgroup. In this update method, and as the number of members of each malware family increases, the representative behavior profiles decrease.

\vspace{0.25cm}

\item {\bf Method 2}: The second update method is {\em union}. The representative behavior profile for each malware family is updated by the union of behavior profiles of members in each subgroup. In this update method, as the number of members of each malware family increases, the representative behavior profiles increase.
\end{enumerate}

We define the similarity score as the intensity with which resources are accessed. Access to resources includes hardware resources (e.g., Call, SMS, Bluetooth, and Camera), system information, and private information (as detailed earlier); we define the similarity score as the weighted sum of the similarity of four behavior factors. The similarity score between the behavior profile of malicious application and a representative behavior profile for each malware family is given by:

\begin{eqnarray} \label{eq:similarity}
S=\sum_{i} w_{i} \cdot BFS_{i} \hspace{0.5cm} where \hspace{0.25cm} \sum_{i} w_{i} = 1
\end{eqnarray}

where $BFS_{i}$ and $w_{i}$ are the similarity and weight of behavior factor $i$, respectively. Similarity of behavior factor (BFS) is composed of four parts: similarity of sending premium-rate SMS (SS), calling premium-rate number (CS), sending sensitive information (SIS), and converting data (CDS). We choose the weight ($w_i$) to be 0.33 for SS, 0.33 for CS, 0.21 for SIS, and 0.13 for CDS---we determined that such settings for weight values are optimal and provide best performance through experiments.

\begin{table*}[htb]
\centering
\renewcommand{\arraystretch}{1.3}
\caption{Similarity metric to apply to each behavior factor.}
\label{Similarity metric}
\begin{footnotesize}
\setlength{\tabcolsep}{12pt}
\begin{tabular}{c c c}
\hline
Behavior factor & Behavior target & Similarity metric \\
\hline
Sending SMS & Premium--rate & Binary (0 or 1) \\
\hline
Calling & Premium--rate & Binary (0 or 1) \\
\hline
Sending sensitive information & \begin{tabular}{@{}c@{}}System information, \\ Private information\end{tabular} & Jaccard index [0, 1] \\
\hline
\multirow{4}{*}{Converting data} & Destination URL &
\begin{tabular}{@{}c@{}}Modified levenshtein \\ distance [0, 1]\end{tabular} \\
\cline{2-3}
& \begin{tabular}{@{}c@{}}Cipher algorithm \\ (DES, AES, Blowfish) \end{tabular} & Binary (0 or 1) \\
\cline{2-3}
& Encoding algorithm (Gzip or not) & Binary (0 or 1) \\
\hline
\end{tabular}
\end{footnotesize}
\end{table*}

Table \ref{Similarity metric} shows similarity metric to apply to each behavior factor, and we compute the similarity score for each behavior factor as follows:

\begin{enumerate}

\item We compute the similarity score for sending premium--rate SMS and calling premium--rate number, as comparing whether a relevant hardware resource is accessed or not. String similarity (e.g., phone number, code number) is less meaningful as a feature except for perfect matching since a difference of one bit yields the same result as with the difference of all bits in this case. Therefore, we give a similarity score of one if they have the same behavior; otherwise, we give a score of zero. Hence, the value of similarity score for both SS and CS is binary.
\item We compute the similarity score for sending sensitive information by applying the Jaccard index. We define the sensitive information as follows (highlighted in Table~\ref{Example} by an example):
	\begin{enumerate}
		\item {\em System information:} IMEI, IMSI, device ID, MCC, MNC, carrier name, device type, device model, OS version.
    		\item {\em Private information:} external storage contents, location, country code, language.
    \end{enumerate}

    We compute the similarity score for converting data (CDS), as the average of the similarity for a destination URL, cipher algorithm, and encoding algorithm. In the case of similarity of a destination URL, we first adopt the longest prefix matching. If a partial matching occurs, we adopt the Levenshtein distance to the residual string except the substring to which the longest prefix matching is used. For example, let A.B.C.D and A.B.E.F be two URLs. In this case, we adopt Levenshtein distance to the residual URLs: C.D and E.F. As for the cipher algorithm and encoding algorithm, we give a similarity score of one if they have used the same algorithm; otherwise, we give a score of zero. The value of similarity score for both SIS and CDS was [0, 1].

\vspace{0.25cm}
\item If a given application does not act maliciously (based on the defined criteria above) except for CDS, we consider that application to be benign.

\end{enumerate}

\section{Performance Evaluation}\label{sec:performance}
In the following we demonstrate the performance and accuracy of Andro-profiler by highlighting aspects of implementation and testing it on various real-world mobile malware samples and families.
\subsection{Implementation}
Our anti-malware system is composed of a mobile device and a remote server; the client application is installed on the mobile device (SKY IM-A690S) running on the Android 2.3.3, and three components---a crawler, repository, and analyzer---were installed on the remote server. The remote server has an Intel(R) Xeon(R) X5660 processor and 4GB of RAM with 32-bit Ubuntu 12.04 LTS operating system; we performed all experiments in a hypervisor-based virtualization environment---VMWare ESXi; \url{http://www.vmware.com/}.

We implemented each component of our anti-malware system with Python high level programming language (as scripts) as follows:

\begin{enumerate}
\item The client component on the mobile device is implemented in the form of an application and communicated with the remote server. The crawler component sent the package name to GooglePlay and downloaded target application. The repository component stored the behavior profile of each application in a database.

\item The analyzer component is composed of the BI, BP, BC, and SM modules. In the following we provide details on each of those modules.
    \begin{enumerate}

    \item The BI module is implemented as python script coupled with Droidbox. The emulator is run on the Android 2.3.4 (level 10). In order to capture the malicious behavior, the BI module executed each application for 60 seconds after the installation process is completed. After capturing integrated system logs of malicious application, the BI module passed those logs to the BP module and restored the emulator to the initial state only for capturing malicious behavior.

    \item The BP module parsed integrated system logs to make the behavior profile of each malware, and stored the behavior profile as a dictionary structure of the Python language for efficient membership test. The parsing rule listed in Table~\ref{Parsing_rule} consists of system call and its arguments---only arguments provided by LKM, and information provided by Droidbox. The parsed behavior profile is encoded in a base-64 format and stored in database.

    \item The BC module categorized malicious application according to the behavioral patterns, and the SM module computed similarity score between behavior profile of malicious application and representative behavior profile for each family. The SM module classified a malicious application into the group with highest similarity score, which is at least 0.85. Whenever a new malware sample is queued into our anti-malware system for inspection, the SM module had continuously updated representative behavior profile according to the pre-chosen update method. 
    \end{enumerate}
\end{enumerate}

\begin{table*}[htb]
\centering
\renewcommand{\arraystretch}{1.3}
\caption{Example of parsing rules for detecting malicious behavior.}
\label{Parsing_rule}
\begin{footnotesize}
\setlength{\tabcolsep}{2pt}
\begin{tabular}{c c c}
\hline
Behavior factor & Parsing rule & Comment \\
\hline
Sending SMS & mms.transaction.SmsReceiverService & SMS \\
\hline
\multirow{2}{*}{Calling} & access(/system/app/Phone.apk $\sim$ )  & \\
& writev(3, OutgoingCallBroadcaster $\sim$ )  & \multirow{-2}{*}{Calling} \\
\hline
\multirow{20}{*}{\minitab[c]{Sending sensitive\\ information}} &
open$($/proc/cpuinfo $\sim$ ),   write(1, Processor $\sim$ )   & CPU Spec. \\
\cline{2-3}
& open(/sdcard $\sim$ ), stat64(/sdcard/ $\sim$ )  & Storage access \\
\cline{2-3}
& stat64$($/system/app/MediaProvider.apk), & \\
& access$($/data/$\sim$/com.android.providers.media/databases), &  \\
& com.android.providers.media.MediaScannerService), & \\
& \minitab[c]{open(/data/dalvik-cache/system$@$app \\ $@$MediaProvider.apk$@$classes.dex)}& \multirow{-4}{*}{Media file} \\
\cline{2-3}
& \{stat64 $|$ open $|$ access\}(/system/app/Contacts.apk), & \\
& \{stat64 $|$ open\} (/data/$\sim$ @Contacts.apk@classes.dex) & \multirow{-2}{*}{Contact information} \\
\cline{2-3}
& \minitab[c]{ $\langle$map$\rangle$ $\sim$ $\{$ NET\_OP $|$ mcc $|$ mnc \} $\sim $ $\langle$$\setminus$map$\rangle$, \\ $\langle$map$\rangle$ $\sim$ $\{$ networkOperator $|$ sim\_operator \} $\sim$ $\langle$$\setminus$map$\rangle$ } & MCC, MNC \\
\cline{2-3}
& $\langle$map$\rangle$ $\sim$ $\{$ affid $|$ did $|$ device\_id $|$ andide \} $\sim$ $\langle$$\setminus$map$\rangle$ & Device ID \\
\cline{2-3}
& $\langle$map$\rangle$ $\sim$ $\{$ osversion $|$ device\_type \} $\sim$ $\langle$$\setminus$map$\rangle$ & OS version \\
\cline{2-3}
& \minitab[c]{$\langle$map$\rangle$ $\sim$ $\{$ manufacturer $|$ phoneModel $|$\\ device\_name $|$ model \} $\sim$ $\langle$$\setminus$map$\rangle$} & Device \\
\cline{2-3}
& $\langle$map$\rangle$ $\sim$ $\{$ network $|$ wifi \} $\sim$ $\langle$$\setminus$map$\rangle$ & Wifi information \\
\cline{2-3}
& $\langle$map$\rangle$ $\sim$ $\{$ carrier $|$ device\_carrier \} $\sim$ $\langle$$\setminus$map$\rangle$ & Carrier \\
\cline{2-3}
& $\langle$map$\rangle$ $\sim$ $\{$ imei $|$ imsi \} $\sim$ $\langle$$\setminus$map$\rangle$ & IMEI, IMSI \\
\cline{2-3}
& $\langle$map$\rangle$ $\sim$ $\{$ longitude $|$ latitude \} $\sim$ $\langle$$\setminus$map$\rangle$ & Location \\
\cline{2-3}
& $\langle$map$\rangle$ $\sim$ $\{$ location $|$ country\_code $|$ locale \} $\sim$ $\langle$$\setminus$map$\rangle$ & Country code \\
\cline{2-3}
& $\langle$map$\rangle$ $\sim$ $\{$ language \} $\sim$ $\langle$$\setminus$map$\rangle$ & Language \\
\hline
\multirow{3}{*}{Converting data}& \minitab[c]{$\{$sendto $|$ OpenNet $|$ SendNet $|$ DataLeak\}\\( $\sim$ Content-Encoding: gzip $\sim$ )}  & Encoding algorithm\\
\cline{2-3}
& \minitab[c]{$\{$sendto $|$ OpenNet $|$ SendNet $|$ DataLeak\}\\( $\sim$ CryptoUsage: $\{$DES$|$AES$|$Blowfish\} $\sim$ )} & Cipher algorithm \\
\hline
\end{tabular}
\end{footnotesize}
\end{table*}

\subsection{Experiment Setup}

For performance evaluation, 643 malware samples consisting of 5 malware families were collected from January 2013 to August 2013 through malware repositories such as virusshare \cite{Virusshare}, contagio \cite{Contagio}, and 8,840 benign samples were collected through GooglePlay for the same periods. In the real world, malware comprises a small fraction of all android apps, so it makes sense to use a larger set of benign samples to mimic the realistic scenario. Duplicated malware samples were eliminated according to SHA 256, and duplicated benign samples were eliminated according to the application package name and SHA 256. We also excluded malware samples diagnosed by fewer than 9 AV vendors included by the VirusTotal dataset \cite{virustotal}. We used textual description of malware produced by F-Secure \cite{F_Secure_desciption}. The details of the dataset we used are in Table~\ref{Mal_sample}.

\begin{table*}[htb]
\centering
\renewcommand{\arraystretch}{1.3}
\caption{Malware samples and benign samples for experiments.}
\label{Mal_sample}
\begin{footnotesize}
\setlength{\tabcolsep}{8pt}
\begin{tabular}{c c c c}
\hline
Category & Family & Quantity & Behavioral characteristics \\
\hline
\multirow{5}{*}{Malware (643)} & AdWo & 401 & Collect the sensitive information \\
\cline{2-4}
& AirPush & 60 & Send SMS \& collect the sensitive information\\
\cline{2-4}
& FakeBattScar  & 44 & Collect the sensitive information \\
\cline{2-4}
& Boxer & 42 & Send SMS \& collect the sensitive information\\
\cline{2-4}
& GinMaster & 96 & Collect the sensitive information \\
\hline
\multirow{2}{*}{ Benign (8,840)} & Application & 7,164 & Normal application \\
\cline{2-4}
& Game & 1,676 & Normal game application \\
\hline
\end{tabular}
\end{footnotesize}
\end{table*}

For the validation of our work, we used 5-fold cross-validation to evaluate the performance in our experiments. The $k$-fold cross-validation is a widely used technique in machine learning. In a nutshell, the method partitions the dataset into $k$ equal size subsets, where each subset is used only once for testing and validation of the training model, and the $k-1$ remaining subsets are used for training the model. This is, a model is built using $k-1$ subsets, and tested using the remaining subset. Then, the subset used in the previous step for testing is used for training, and a subset in the $k-1$ sets previously not used for testing is used for testing. The process is repeated $k$ times by alternating the testing set, and the results are averaged over the runs.

\subsection{Comparing Different Methods}
To the best of our knowledge, the closest approach in the literature to Andro-profiler is Crowdroid \cite{Burguera2011}. Crowdroid monitors invoked system calls and makes frequency table of system calls at the client side. Crowdroid identifies malicious behavior and detects malware utilizing the $K$-means algorithm at the server side. For the completeness of our work, we provide an experimental comparison between Andro-profiler and Crowdroid. To conduct more fair performance evaluation and comparison, we make both systems work in a similar context and using similar settings: we modify Crowdroid to hook all system calls invoked during the execution processes, including the installation phase.

\subsection{Selection of Weight for Behavior Factors}

Andro-profiler needs to select appropriate weights ($w_{i}$) in order to guarantee the best performance. However, we cannot obtain a unique solution of equation (\ref{eq:similarity}) analytically, because there are only two equations given in order to compute values of four variables, which means that we cannot obtain an optimal solution of equation (\ref{eq:similarity}). We might obtain local optimum values of equation (\ref{eq:similarity}) through simple numerical approach (iterative method) as follows. First, we setup initial values of weight by solving arithmetic mean of them. We apply those values to the equation (\ref{eq:similarity}), then evaluate the classification capability. Next, we increase the weight of SS and CS, and decrease the weight of SIS and CDS. We then apply those values to the equation (\ref{eq:similarity}), and conduct the evaluation of classification capability iteratively. The reason we determine that the weight of CDS is smaller than other factors is as follows. First, if a client cannot connect to the remote server, a malware sample does not need to convert format of data for transmitting sensitive information. Second, benign applications also need an encoding algorithm for efficient transmission and cipher algorithm for secure communication. We adjust the weight of SIS in order to maximize the effect of calling and sending premium-rate SMS.

We proceed with the iterative steps until the tendency of classification accuracy is changed. We believe that our system reaches a local optimum at that point. Table \ref{Weighted_Val} shows that the results of simple numerical approach according to weight change.
We choose the value of the weight ($w_{i}$) to be 0.33 for SS, 0.33 for CS, 0.21 for SIS, and 0.13 for CDS, since it provides a good performance that matches close to the ground truth.

\begin{table}[htb]
\centering
\renewcommand{\arraystretch}{1.3}
\caption{The classification accuracy and the number of cluster according to changes of weight (e.g., Method 1). The number of clusters means that the number of groups that malware/benign samples are classified into. Bold text means that the tendency of classification accuracy is
changed. At this point, we believe, our system reaches a local optimum for the best performance.}
\label{Weighted_Val}
{\tiny
\setlength{\tabcolsep}{0pt}
\begin{tabular}{c c c c c c c c}
\hline
\multirow{2}{*}{No} & \multicolumn{4}{c}{Weight of behavior factor} & \multicolumn{2}{c}{Number of clusters} & \multirow{2}{*}{Accuracy} \\
\cline{2-7}
 & SS & CS & SIS & CDS & Malware & Benign & \\
\hline
1 & 0.25 & 0.25 & 0.25 & 0.25 & 8 & 4 & 0.98 \\
\hline
2 & 0.27 & 0.27 & 0.24 & 0.22 & 6 & 2 & 0.98 \\
\hline
3 & 0.29 & 0.29 & 0.23 & 0.19 & 6 & 2 & 0.98 \\
\hline
4 & 0.31 & 0.31 & 0.22 & 0.16 & 6 & 2 & 0.98 \\
\hline
5 & \textbf{0.33} & \textbf{0.33} & \textbf{0.21} & \textbf{0.13} & \textbf{6} & \textbf{1} & \textbf{0.98} \\
\hline
6 & \textbf{0.35} & \textbf{0.35} & \textbf{0.20} & \textbf{0.10} & \textbf{6} & \textbf{1} & \textbf{0.98} \\
\hline

\end{tabular}
}
\end{table}

\subsection{Experiment Results and Analysis}
Our performance evaluation focuses on the effectiveness of malware classification, discriminatory ability between malware and benign applications, and the efficiency of malware classification. We demonstrate that our system performs well in detecting and classifying malware families. We used the accuracy as the performance metric, since the metric for performance evaluation must focus on the predictive capability of the model. We measured the accuracy as the total number of the hits of the classifier divided by the number of instances in the whole dataset. The performance of malware classification model is determined by how well the model detects and classifies various pieces of malware. We measured the accuracy as the total number of the hits of the classifier divided by the number of instances in the whole dataset.  Moreover we used the Receiver Operating Characteristic (ROC) curve as the method for comparing classification models. To compare the ROC performance of classifiers intuitively, we calculated the AUC (area under the curve; also known as the integral) of each classifier, since the AUC represents the ROC performance in a single scalar value \cite{fawcett2006}.

\subsubsection{Effectiveness of Malware Classification}

First, we demonstrate that our proposed method provides effective metric to detect and classify malware families.
Table \ref{Sim_Mals} presents the result of similarity comparison with the representative profile of each malware family and benign applications.
Despite that {\tt Boxer} sends premium-rate SMS according to anti-virus (AV) analysis report, our emulator-based approach fails to capture sending premium-rate SMS due to connection error; our method only captures sending sensitive information. However, our system performs well in classify all malware including {\tt Boxer}.
Since the difference of similarity score among all malware is smaller than the threshold (0.85), that can be good metric for detect and classify malware. 
The difference of similarity score for {\tt AirPush} is much larger than the others, because {\tt AirPush} sends premium-rate SMS and sends sensitive information while the other malware families send sensitive information. Since benign applications do not act maliciously, it is natural that the difference of similarity score between malware and benign applications is large based on the metrics and features utilized for computing the behavior profile.

\begin{table}[h!]
\centering
\renewcommand{\arraystretch}{1.3}
\caption{The similarity comparison with representative behavior profile of each malware family and benign.}
\label{Sim_Mals}
{\tiny
\setlength{\tabcolsep}{1pt}
\begin{tabular}{c c c c c c c}
\hline
Similarity & AdWo & AirPush & Boxer & FakeBattScar & GinMaster \\\hline
AdWo         & -    & 0.37 & 0.70 & 0.70 & 0.70 \\\hline
AirPush      & 0.37 & -    & 0.46 & 0.46 & 0.50 \\\hline
Boxer        & 0.70 & 0.46 & -    & 0.79 & 0.79 \\\hline
FakeBattScar & 0.70 & 0.46 & 0.79 & -    & 0.79 \\\hline
GinMaster    & 0.70 & 0.50 & 0.79 & 0.79 & - \\\hline
Benign       & 0.04 & 0.13 & 0.13 & 0.13 & 0.13 \\\hline
\end{tabular}
}
\end{table}

Next, Table \ref{Clas_Accuracy_AUC_1} shows that Andro-profiler performs well in classifying malware families with $100\%$ classification accuracy on average, regardless of the update method. Furthermore, Andro-profiler is shown to outperform Crowdroid, which gives an average classification accuracy of $49\%$.
Some factors may have affected that Crowdroid underperforms the Andro-profiler. Since invoked system calls among malware families are similar to each other, Crowdroid limits to classify malware families; malware families mainly call out system calls (e.g., read(), close(), open(), write(), recvmsg()). Since FakeBattScar calls out more system calls (e.g., open(), close()) than others and Adwo calls out system call of read() constantly, two malware families can be classified well.
Furthermore, Andro-profiler gives $47\%$ performance improvement advantage over Crowdroid in terms of the AUC. 
In the case of method 1, our system clusters {\tt Airpush} samples into two groups. We conducted a deep analysis to understand the reason method 1 of our system clustered such samples into two groups, and found that almost half of Airpush samples sent premium-rate SMS and collected sensitive information (e.g., IMEI, Android version, location information, and carrier), whereas the other half only collected sensitive information. To this end, we found that our system identified malicious behavior and classified malware according to behavior patterns of malware families.

\begin{table}[h!]
\centering
\renewcommand{\arraystretch}{1.3}
\caption{Classification performance for 643 malware. Bold text means that Andro-profiler outperforms Crowdroid in classifying malware families.}
\label{Clas_Accuracy_AUC_1}
{\tiny
\setlength{\tabcolsep}{0pt}
\begin{tabular}{c c c c c c c c }
\hline
\multicolumn{2}{c}{ \multirow{3}{*}{Category} } & \multicolumn{3}{c}{Accuracy} & \multicolumn{3}{c}{AUC} \\
\cline{3-8}
& & Method 1 & Method 2 & Crowdroid & Method 1 & Method 2 & Crowdroid \\
\hline
\multirow{5}{*}{Malware} &AdWo & \textbf{1.00} & \textbf{1.00} & 0.83 & \textbf{1.00} & \textbf{1.00} & 0.73 \\
\cline{2-8}
&AirPush & \textbf{1.00} & \textbf{1.00} & 0.02 & \textbf{1.00} & \textbf{1.00} & 0.51 \\
\cline{2-8}
&Boxer & \textbf{1.00} & \textbf{1.00} & 0.37 & \textbf{1.00} & \textbf{1.00} & 0.63 \\
\cline{2-8}
&FakeBattScar & \textbf{1.00} & \textbf{1.00} & \textbf{1.00} & \textbf{1.00} & \textbf{1.00} & \textbf{1.00} \\
\cline{2-8}
&GinMaster & \textbf{1.00} & \textbf{1.00}  & 0.22 & \textbf{1.00} & \textbf{1.00}  & 0.54 \\
\hline
\multicolumn{2}{c}{Average} & \textbf{1.00} & \textbf{1.00} & 0.49 & \textbf{1.00} & \textbf{1.00} & 0.68 \\
\hline
\end{tabular}
}
\end{table}

\subsubsection{Discriminatory Ability Between Malware and Benign}

When designing an anti-malware system, one important factor which we should also consider is its discriminatory ability between malware and benign applications. Anti-malware systems must detect malware with small errors in terms of false positive and false negative. We believe that it is more important for an anti-malware system to detect malware with small false negative than false positive. However, for commercial reasons, one may think the opposite: users can be bothered if their benign applications are misclassified as malware. Since our method classified five malware families and benign applications, and for avoiding the ambiguity of interpretation, we do not adopt false positive rate and false negative rate as a performance metric. Instead, we used the accuracy and AUC as the performance metric. Table~\ref{Clas_Accuracy_AUC_2} shows that Andro-profiler performs well in detecting and classifying malware families with $98\%$ classification accuracy on average, regardless of the update method, while Crowdroid detects malware families with $90\%$ classification accuracy on average.
Some factors may have affected that Crowdroid underperforms the Andro-profiler. Since invoked system calls between malware and benign samples are similar to each other, Crowdroid limits to detect and classify malware families; all samples mainly call out system calls (e.g., read(), close(), open(), write(), recvmsg()). Among these, FakeBattScar calls out more system calls (e.g., open(), close()) than others and other malware families have similar call frequencies to benign samples, then malware families except for FakeBattScar cannot be detected and classified well.
Our proposed methods also outperform Crowdroid by improving its AUC by about $90\%$.

\begin{table}[htb]
\centering
\renewcommand{\arraystretch}{1.3}
\caption{Classification performance for 643 malware and 8,840 benign samples. Bold text means that Andro-profiler outperforms Crowdroid in detecting malware and classifying malware families.}
\label{Clas_Accuracy_AUC_2}
{\tiny
\setlength{\tabcolsep}{0pt}
\begin{tabular}{c c c c c c c c }
\hline
\multicolumn{2}{c}{ \multirow{3}{*}{Category} } & \multicolumn{3}{c}{Accuracy} & \multicolumn{3}{c}{AUC} \\
\cline{3-8}
& & Method 1 & Method 2 & Crowdroid & Method 1 & Method 2 & Crowdroid \\
\hline
\multirow{5}{*}{Malware} &AdWo & \textbf{1.00} & \textbf{1.00} & 0.01 & \textbf{1.00} & \textbf{1.00} & 0.49 \\
\cline{2-8}
&AirPush & \textbf{1.00} & \textbf{1.00} & 0.00 & \textbf{1.00} & \textbf{1.00} & 0.50 \\
\cline{2-8}
&Boxer & \textbf{1.00} & \textbf{1.00} & 0.00 & \textbf{1.00} & \textbf{1.00} & 0.50 \\
\cline{2-8}
&FakeBattScar & \textbf{1.00} & \textbf{1.00} & \textbf{1.00} & \textbf{1.00} & \textbf{1.00}  & \textbf{1.00} \\
\cline{2-8}
&GinMaster & \textbf{1.00} & \textbf{1.00} & 0.00 & \textbf{1.00} & \textbf{1.00}  & 0.49 \\
\hline
\multicolumn{2}{c}{Benign}  & \textbf{0.97} & \textbf{0.97} & 0.96 & \textbf{0.99}  & \textbf{0.99}  & 0.52 \\
\hline
\multicolumn{2}{c}{Average} & \textbf{0.98} & \textbf{0.98}  & 0.90 & \textbf{0.99} & \textbf{0.99}  & 0.52 \\
\hline
\end{tabular}
}
\end{table}

Andro-profiler misclassified 225 benign samples as malware. We conducted a deep analysis to understand the high false positives with Andro-profiler. Interestingly, we found that some benign samples collected user's sensitive information, which we defined as a trigger for classifying malicious applications (e.g., IMEI, device ID, UUID, latitude, and longitude). To understand whether other anti-malware systems and scanners considered those benign applications as malware or not, we uploaded those suspected GooglePlay samples to VirusTotal and checked scanning results of various anti-virus vendors. As a result, we found that 110 out of the suspicious benign samples (accounting for about $49\%$) were diagnosed as malware. The high rate of misclassification of benign applications is, however, understandable given various potential reasons for such infiltration of gray area applications into the market place~\cite{bouncer}.

\subsubsection{Effectiveness of Detecting Zero-Day Malware}

We demonstrate the effectiveness of detecting zero-day malware detection.
We define an application as a zero-day malware if it has malicious behavior and it cannot be detected by AV vendors.
In order to verify that we had appropriately detected zero-day malware, we made 91 variant samples consisting of Adwo and AirPush families by leveraging ADAM \cite{Adam2013}. All samples used as the base application for the variants are among the ones which are used in the previous experiments, and detected by VirusTotal as malware samples. After creating the variants, we uploaded them (as samples) to the VirusTotal, and checked scanning results of various anti-virus (AV) vendors such as F-Secure, Kaspersky, ClamAV, and Avast.
We noted that none of the submitted samples is reported as a malware when we carried out our experiment.
As a result of our experiment using Andro-profiler, we found that it performed well in detecting all of the variant malware samples with $100\%$ classification accuracy on average, regardless of the update method.

\subsubsection{Efficiency of Malware Classification}

Our proposed system only takes 55 seconds/MB for classifying each malware; we exclude setup time for analysis such as booting time of emulator. The majority of this time is spent in making the behavior profile; it takes only 0.2 seconds on average to classify malware into each family.

While the performance of our system is operationally reasonable, our system is scalable both horizontally and vertically by design. Horizontally, and given that our server side components are run in a virtual environment, one can fork multiple servers by utilizing multiple virtual machines that exploit the multi-core nature of today's commodity computers. Vertically, our system can benefit from being developed in a lower level language, such as the C language, which would make the classification process run faster.

\section{Limitations}

Andro-profiler has a few limitations for detecting and classifying malware, since our proposed method uses integrated system logs as a feature vector and employs dynamic analysis techniques to capture malware's behavior. First, it is difficult for our system to analyze malware that are executed only under given conditions (e.g., SDK version, cellular network connection status, time, or place). However, this shortcoming is addressable by having various platforms tailored with various settings, as used for traditional malware in~\cite{mohaisen2013amal}. It is also impossible for our system to analyze malware embedding anti-malware analysis techniques. Second, our emulator-based anti-malware system is dependent on SDK version of emulator, so our approach has limitation on analyzing malicious behavior related to privilege escalation. However, those are common drawbacks of dynamic analysis method or emulator-based detection method and addressed in the literature at some expense.

Finally, our approach analyzes malware on an emulator without interaction between human and device: autonomous installation and execution. When a malware behave upon an update or by utilizing a drive-by download attack \cite{Yajin2012}, our approach is limited in reacting to such malware. However, autonomous installation and execution is an inevitable procedure for automation of dynamic analysis. Depending on the number of malware samples to be analyzed, one can adopt manual human interactions to analyze malware samples and vet the outcomes of the automatic classification procedure, as used in~\cite{mohaisen2013amal}.

\section{Conclusion and Future Work}\label{conclusion}
In this paper, we have presented  Andro-profiler, an anti-malware system based on behavior profiling. Using Andro-profiler, we classified malware by exploiting the behavior profiling extracted from integrated system logs, which are implicitly equivalent to distinct behavior characteristics. Our behavior profiling is simple and relatively easy to understand, whereas Andro-profiler is capable of distinguishing benign and malicious applications, and malicious applications into families. Furthermore, Andro-profiler is capable of detecting zero-day threats, which are missed by antivirus scanners.

Our experiments demonstrate that Andro-profiler performs well in detecting and classifying malware families with over $98\%$ classification accuracy on average regardless of update method while Crowdroid, a closely related work from the literature, performs under $90\%$ classification accuracy on average. Our experiment results indicate that it takes 55 seconds/MB to analyze a malware on average, with a lot of opportunities for improvements on scalability. Our system hence enables AV vendors to react to many species of malicious samples by classifying and matching these with previous ones effectively and efficiently.

There are several directions that we will pursue in the future. First, we would like to augment our system to not only rely on dynamic and behavioral features, but also static features that are easy to obtain from the applications at scale. Furthermore, we will explore scalability issues associated with our system by implementing some of the guidelines noted in subsection \emph{Efficiency of Malware Classification}.


%

%

\section{Acknowledgements}
This research is supported by the MSIP (Ministry of Science, ICT and Future Planning), Korea, under the ITRC (Information Technology Research Center) support program (NIPA-2014-H0301-14-1004) supervised by the NIPA(National IT Industry Promotion Agency). In addition, this work is also supported by the ICT R\&D Program of MSIP/IITP. [14-912-06-002, The Development of Script-based Cyber Attack Protection Technology]. A two-page abstract on this work appeared in \cite{JangYWK14}. The work proposed in this paper significantly enhances the prior work, technically and content-wise, including the motivation, related-work, design, and evaluation.

\end{document}